# Quantum relays and noise suppression using linear optics

B. C. Jacobs, T. B. Pittman, and J. D. Franson

Johns Hopkins University

Applied Physics Laboratory

Laurel, MD 20723

Abstract:

Probabilistic quantum non-demolition (QND) measurements can be performed using linear optics and post-selection. Here we show how QND devices of this kind can be used in a straightforward way to implement a quantum relay, which is capable of extending the range of a quantum cryptography system by suppressing the effects of detector noise. Unlike a quantum repeater, a quantum relay system does not require entanglement purification or the ability to store photons.

Probabilistic quantum logic operations can be implemented using linear optical elements, additional photons (ancilla), and post-selection [1,2]. We have proposed [3] and experimentally demonstrated [4,5] several logic devices of this kind that succeed with an ideal probability of ½, while the probability of success can be made arbitrarily high using larger numbers of ancilla [6]. Here we show that probabilistic quantum non-demolition (QND) measurements [7,8] based on these techniques can be used to implement a quantum relay that can extend the throughput and maximum range of a quantum cryptography system by suppressing the noise due to detector dark counts. Unlike a quantum repeater [9], a quantum relay does not require entanglement purification [10] or the ability to store photons.

The specific QND implementation that we present here is a modification of a probabilistic quantum encoder circuit [3]; however, the results of our relay analysis are applicable to other QND implementations as well [8]. As shown in Fig. 1, the encoder circuit conditionally encodes (copies) the state of an input qubit into two output qubits. As will be shown below, the addition of a second detector can be used to signal the presence of an input photon while the polarization state of the input qubit is transferred into the remaining output. As Kok, Lee, and Dowling [8] recently pointed out, quantum teleportation [11] can be used to implement QND measurements; accordingly, Fig. 1 can be seen to be a teleportation-based QND device using the Bell-state measurement approach of Pan and Zeilinger [12].

We begin by describing the operation of the quantum encoder and its modification to perform QND measurements. The concept of a quantum relay is then introduced, in which each segment of a communications channel conditionally passes (relays) a qubit on to the next segment of the communications channel provided that a QND measurement has verified that a photon is actually present. This does not avoid the exponential loss of signal in an optical fiber, but the limiting effects of detector dark counts on error correction and privacy amplification can be essentially eliminated. The performance of such a quantum relay is then analyzed in terms of its potential for increased throughput and operating range.

As shown in Fig. 1, the QND measurement of interest here is implemented using polarizing beam splitters. Its intended function is to produce a classical output signal if one and only one photon is present in the input, while transferring the polarization state of the incident qubit to the output mode. The output of the QND measurement is known to be correct whenever one and only one photon is detected in each detector assembly, which occurs with a probability of ½.

The notation used here is the same as in our earlier paper on quantum logic operations [3]. Qubit values 0 and 1 ($|0\rangle$ and $|1\rangle$) are represented by the horizontal and vertical polarization modes ($|H\rangle$ and $|V\rangle$) of single photons. The input qubit is assumed to be in an arbitrary superposition state given by $|\psi\rangle_0 \equiv \alpha|H\rangle_0 + \beta|V\rangle_0$, where $\alpha$ and $\beta$ are complex coefficients.

The intended function of the encoder circuit is to perform the transformation

$$\alpha|H\rangle_0 + \beta|V\rangle_0 \rightarrow \alpha|H\rangle_1|H\rangle_2 + \beta|V\rangle_1|V\rangle_2 \qquad (1)$$

As shown in Fig. 1, this can be accomplished by using a polarizing beam splitter to mix the input mode (0) with one photon (a) that is part of a pair of entangled ancilla photons emitted by $\phi^+$ into modes a and 1. The output modes (1 and 2) are conditionally accepted if a polarization-sensitive detector package $D_b$ (shown in the inset of Fig. 1) records one-and-only-one (1AO1) event. The detector package $D_b$ consists of a polarizing beam splitter rotated 45° from the HV basis, followed by two ordinary single-photon detectors; the axes of the rotated basis will be referred to as F and S:

$$F \equiv \frac{1}{\sqrt{2}}[H+V] \tag{2}$$

$$S \equiv \frac{1}{\sqrt{2}}[H-V] \tag{3}$$

Here the ket notation has been dropped for compactness. The 1AO1 condition signals the successful projection of the combined state onto the desired output state, which is the origin of the nonlinearity required for logic operations.

The entangled ancilla photons are created in a Bell state of the form: $\phi^+_{a1} \equiv \frac{1}{\sqrt{2}}[H_a H_1 + V_a V_1]$. The state of the system after the polarizing beam splitter can be shown to be

$$\psi_{12b} = \frac{1}{\sqrt{2}}[\alpha H_1 H_2 H_b + \beta V_1 V_2 V_b + \alpha V_1 H_2 V_2 + \beta H_1 V_b H_b], \tag{4}$$

The last two terms in Eq. (4) correspond to zero or two photons going to detector package $D_b$, and these terms are therefore projected out of the accepted state (with a probability of 1/2) by the 1AO1 condition. The usefulness of the projected state $\psi_P$ becomes apparent when it is renormalized and expressed in the FS basis:

$$\psi_P = \frac{1}{\sqrt{2}}[(\alpha H_1 H_2 + \beta V_1 V_2)F_b + (\alpha H_1 H_2 - \beta V_1 V_2)S_b] \tag{5}$$

It can be seen from Eq. (5) that the quantum encoder performs the desired logic operation whenever 1AO1 photon is found in the $F_b$ channel. In addition, the feed-forward quantum control methods that we have recently demonstrated [5] can be used to obtain the desired output for $S_b$ detection events by reversing the relative sign of the $\alpha$ and $\beta$ terms.

The encoder circuit can be converted to a QND measurement device by adding a second polarization-sensitive detector package $D_2$ that is identical to $D_b$, but located in path 2. If one and only one photon is detected in both detector packages, the projected state of the system can be shown to be

$$\psi_{P2} = \frac{1}{2}\left[\alpha H_1(F_2 F_b + S_2 F_b + F_2 S_b + S_2 S_b) + \beta V_1(F_2 F_b - S_2 F_b - F_2 S_b + S_2 S_b)\right] \quad (6)$$

It can be seen that the output state in mode 1 is identical to the input state under these conditions, provided that feed-forward techniques [5] are used to reverse the relative sign of the $\alpha$ and $\beta$ terms for some of the combinations of $D_b$ and $D_2$ detection events, such as $S_2 F_b$, for example. The complete circuit probabilistically implements a QND measurement on a photon in the sense that a classical signal is generated only when an input photon is present without affecting its state of polarization. This occurs with a probability of ½ assuming ideal hardware; the effects of detector noise will be considered below.

The reason for placing the second detector package in mode 2 instead of mode 1 can be seen by considering the operation of the device when there is no photon present in the input mode. In that case the joint 1AO1 condition for both detectors cannot be fulfilled because only one of the detection packages receives an ancilla photon. If $D_2$ were moved to output mode 1 instead, then the device could produce a false gate signal when no photon is present in the input because the two ancilla could trigger both detector packages in that case.

Having described a specific method for making quantum non-demolition measurements with a success probability of 1/2, we now focus on a potential application of this type of device in a quantum communications system. The maximum range of current fiber-based quantum cryptography systems is limited by the loss of photons as they propagate through an optical fiber combined with the dark counts in the detectors. Error correction and privacy amplification [13] become increasingly inefficient as the number of remaining photons becomes comparable to the detector dark count, at which point the effective throughput of the system rapidly drops to zero. The range can be extended using quantum repeaters [9] based on entanglement swapping [14] or quantum teleportation [11], but both of these methods require entanglement purification [10] and the ability to store photons for an appreciable time. In contrast, the quantum relay described below can increase the range and total throughput (after error correction and privacy amplification) without the need for entanglement purification or photon storage because all of the required qubit manipulations are local, i.e. self-contained in the relays. Losses in the fiber still occur, but the effects of detector dark counts are suppressed using QND measurements, thereby greatly increasing the efficiency of the privacy amplification and error correction protocols. This scheme is somewhat similar to the notion of event-ready detection [15]; however, we show that by distributing the relays throughout the channel the impact of detector noise, both in the relays and at the receiver, can be made negligible.

The implementation of a quantum relay system using QND measurements is illustrated in Fig. 2. Each relay $R_i$ performs a quantum non-demolition measurement to determine if a photon is present or if it has been lost in transmission through the fiber up to that point. If the photon is still present, a classical gate signal indicating that fact is sent on to the next relay along with the photon itself. If a photon is not detected beyond some point in the transmission line, the gating information is used to ignore that event and not accept any output from the detectors in the receiver. As a result, the dark count rate in the detector will be greatly reduced and the signal to noise ratio S (number of true photon detection events divided by the number of spurious detection events) will be increased compared to its value without any relays. We refer to this system as a quantum relay because each node in the system conditionally passes (relays) a qubit on to the next node, provided a photon was found to be present.

It is obviously important to include the effects of detector dark counts in the relay elements themselves as well as the probability of ½ for the successful operation of the QND measurements. In fact, one might suspect that the relays would only make the situation worse when these factors are taken into account. However, any spurious photons generated by the relays will be attenuated exponentially as they propagate through the fiber. As long as the relay elements are sufficiently far from the receiver, this attenuation will cause the contribution from spurious relay photons to be much smaller than the dark count in the receiver. In the same way, the factor of ½ loss associated with the probabilistic QND measurements can be much smaller than the inefficiency in error correction and privacy amplification that would have occurred without the signal-to-noise improvement from a quantum relay.

An ideal relay element can be viewed as implementing the following transformation on the input density matrix

$$P_1|\theta\rangle\langle\theta| + P_0|0\rangle\langle 0| \rightarrow \frac{P_1}{2}|\theta\rangle\langle\theta| + (1-\frac{P_1}{2})|\varnothing\rangle\langle\varnothing| \qquad (7)$$

Here $P_1$ is the probability that a single photon in the polarization state $|\theta\rangle$ is present, $P_0$ denotes the probability that no photon was present, and the state $|\varnothing\rangle$ represents a situation in which the absence of a gate signal indicates that no photon was present. (The QND measurement also rejects events in which there was more than one photon in the input channel and Eq. (7) could be generalized accordingly.) If dark counts in the relay detectors are included, then the effects of a single relay can be described by

$$P_1|\theta\rangle\langle\theta| + P_0|0\rangle\langle 0| \rightarrow \eta P_1|\theta\rangle\langle\theta| + P_d I + (1-\eta P_1 - 2P_d)|\varnothing\rangle\langle\varnothing| \qquad (8)$$

Here $\eta$ is a reduced efficiency close to ½ (assuming heralded pairs of ancilla photons), I is the identity matrix, and $P_d$ is the probability of a dark count in one of the QND detectors during the processing time of a single qubit. The use of the identity matrix in Eq. (8) reflects the fact that the spurious photons emitted as a result of detector dark counts in the relays have random polarizations. Since $P_d$ is typically very small (~$10^{-5}$ for a 10 MHz system using commercial single-photon counting modules), we only need to consider the probability of a single dark count event occurring in one of the four detectors. Furthermore, since the 1AO1 detection condition correctly excludes half of the dark count events (because they occur in the same package as the ancilla detection), the probability of a relay error in Eq. (8) is approximately $2P_d$ even though four detectors are used in each QND device. This probability of error obviously depends on the specific

QND measurement device used, and can be generalized for other implementations accordingly.

Secure communications in a quantum cryptography system is only guaranteed if the quantum bit error rate $Q_B$ is below the error rate that would be produced by an eavesdropper, which is 25% for an ideal BB84 [16] implementation. Since the maximum range of current optical fiber systems is primarily determined by the impact of exponential photon losses and detector noise on $Q_B$, we will assume an otherwise perfect system, i.e. no optical misalignments or background light. A typical BB84 receiver [17] utilizes two detectors, so that the probability $P_n$ of a noise event in the receiver is roughly twice the detector dark count probability (i.e. $P_n \sim 2 P_d$) in the limit of small $P_d$. (For simplicity, we assume that all of the detectors have the same dark count.) Under these assumptions, the quantum bit error rate for a quantum cryptography system with no quantum relays is given by

$$Q_B = \frac{\frac{1}{2}P_n}{P_n + P_s} = \frac{1}{2(1+S)} \qquad (9)$$

Here $P_s$ is the probability of a signal photon detection and $S \equiv P_s/P_n$. The factor of ½ in Eq. (9) is due to the fact that half of the dark count events accidentally give the correct result. Including the exponential attenuation in the fiber, the signal to noise ratio of an otherwise ideal cryptography system is

$$S_0 = \tfrac{1}{2} P_d^{-1} e^{-\alpha x} \qquad (10)$$

Here $x$ is the transmission distance and $\alpha$ is the fiber attenuation parameter, which is ~0.05/km for a typical optical fiber loss of 0.2dB/km. From Eq. (9) we see that the maximum $Q_B$ threshold of 25% corresponds to a minimum signal to noise ratio of $S=1$.

We now consider the potential improvement in the signal to noise ratio if a single quantum relay is added to the system at a distance $x_1$ from the transmitter and $x_2$ from the receiver, so that $x = x_1 + x_2$. The impact of the relay on the cryptography signal is a straightforward reduction in $P_s$ due to the efficiency of the relay, i.e. $P_s \to \eta e^{-\alpha x}$. Although the relay is designed to reduce noise by gating detector events in the cryptography receiver, it also adds noise in the form of randomly polarized photons. Since these added noise photons must propagate through the channel to reach the receiver, the total noise probability is given by

$$P_n = 2 P_G P_d + P_R e^{-\alpha x_2} \tag{11}$$

Here $P_G$ is the probability that a gate signal will be produced by the relay and $P_R = 2P_d$ is the probability that a spurious output photon will be produced in the relay itself. The first term in Eq. (11) represents the receiver dark count reduction by a factor of $P_G$, while the second term corresponds to the spurious relay photons after attenuation in the fiber. Including the attenuation of signal photons in propagating to the relay gives $P_G = \eta e^{-\alpha x_1}$ (events with multiple dark counts have been neglected), which reduces Eq. (11) to

$$P_n = 2 P_d \left[ \eta e^{-\alpha x_1} + e^{-\alpha x_2} \right]. \tag{12}$$

The optimal single-relay position can be found by minimizing $P_n$, which gives $x_1 = \tfrac{1}{2}\left(x + \alpha^{-1} Log_e[\eta]\right)$; the optimal location is shifted somewhat towards the transmitter (the logarithm is negative). Inserting these values of $x_1$ and $x_2$ into the above equations gives a signal to noise ratio $S_1$ that is exponentially better than that from the cryptography system alone:

$$S_1 = S_0 \left[ \tfrac{1}{2} \eta^{\tfrac{1}{2}} e^{\tfrac{\alpha x}{2}} \right]. \tag{13}$$

Although this result may seem surprising, it can be understood from the fact that the attenuation of the signal before the relay reduces the probability of a gate signal, which in turn reduces the probability of a spurious count in the receiver, while attenuation after the relay reduces the effects of spurious photons generated in the relay.

A similar analysis assuming N relay elements optimally distributed throughout the channel results in a maximum signal to noise ratio $S_N$ given by

$$S_N = S_0 \left[ \tfrac{1}{N+1} \eta^{\frac{N}{N+1}} e^{\alpha x \frac{N}{N+1}} \right] \qquad (14)$$

The signal to noise ratio is not overly sensitive to the relay efficiency, even for a large number of elements, because the signal and noise are equally reduced by all but one relay. The optimum placement of the relays is uniform throughout the channel with the exception of the last relay (nearest the receiver), whose optimal location is given by $\tfrac{1}{N+1}\left(x - N\alpha^{-1} Log_e[\eta]\right)$.

For comparison, we have performed an exact numerical simulation of a quantum cryptography system augmented with quantum relays. This simulation includes the possibility of multiple dark counts that was neglected in the equations above. The results for several values of N are shown in Fig. 3, which indicates good agreement with the approximate analytic results presented above. This plot also suggests that a single quantum relay could approximately double the range of a quantum cryptography system.

The potential range enhancement of a relay system can be estimated by considering a system operated at a fixed signal to noise ratio. By solving Eqs. (10) & (14) for the maximum range at a given value of S, we calculate the range enhancement ratio R to be

$$R \equiv \frac{R_N}{R_0} \approx (N+1)\frac{Log_e[(N+1)\eta^{-N/(N+1)}P_d S]}{Log_e[P_d S]} \quad (15)$$

Although the range enhancement ratio R can be very large, it should be kept in mind that a quantum relay does not avoid the loss in signal due to attenuation. For example, at a range where α$x$=25, the bit rate of a perfect system operating at 100GHz would be ~1 bit per second.

The potential advantage of a quantum relay system can be put in perspective by calculating the overall throughput of a quantum cryptography system including the effects of error correction and privacy amplification, for which we used the results of Lutkenhaus [18]. Fig. 4 shows the results of such an analysis for up to three relays, where the normalized throughput $T_n$ is defined as the total throughput divided by the signal rate after attenuation but prior to error correction or privacy amplification. For a quantum cryptography system with no relays (N=0), the figure clearly shows the dramatic degradation in throughput as the signal to noise ratio approaches 1, which provides an upper bound on the achievable range in that case. It can be seen that the use of quantum relays can postpone this sharp transition, although it should be emphasized once again that losses in the fiber still occur and that the throughput still decreases exponentially as a result.

In summary, we have described the use of probabilistic QND measurements to implement a quantum relay that can extend the range and throughput of a quantum cryptography system despite the non-deterministic nature of the devices involved. The QND measurements described here can be implemented using polarizing beam splitters, post selection, and feed-forward quantum control techniques, with an ideal efficiency of ½. They can be viewed either as a modification of a quantum encoder [3] or as a new application of quantum teleportation [8]. Unlike a quantum repeater [9], a quantum relay does not require entanglement purification or the ability to store photons. On the other hand, the practical applications of a quantum relay are limited by the fact that it does not correct for decoherence of the qubits or the exponential loss of photons in an optical fiber. (Decoherence in quantum cryptography systems is generally quite low and fidelities in excess of 0.99 can be obtained using classical feedback techniques [17].) As recently noted by Kok, Williams, and Dowling [19], the probabilistic CNOT gates that we have described elsewhere [3] could also be used to implement a quantum repeater, which would be more challenging but could, at least in principle, compensate for decoherence and loss as well.

This work was supported by the Office of Naval Research and by IR&D funds.

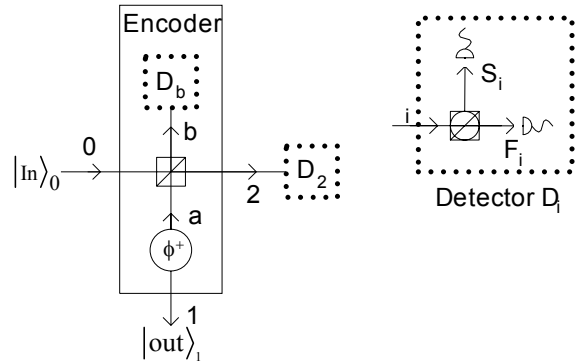

FIG. 1. An implementation of a probabilistic QND measurement of a qubit input in mode 0 using a probabilistic quantum encoder [3] followed by a polarization-sensitive detector in output mode 2. The inset shows the details of the polarization-sensitive detector packages, each of which consists of a polarizing beam splitter rotated through a 45° angle (into the FS basis) followed by two ordinary single-photon detectors. The device can be viewed either as a modification of a quantum encoder circuit [3] or as a new application of quantum teleportation [8,12].

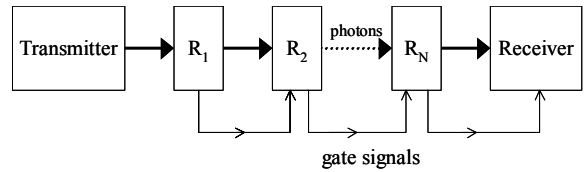

FIG. 2. A quantum relay system in which each relay $R_i$ conditionally passes (relays) a qubit and a gate signal on to the next element provided a QND measurement indicated that a photon was actually present. The quantum relay suppresses the effects of dark counts in the receiver detectors via the gate signal, while spurious photons generated by dark counts in the relays themselves are exponentially attenuated by the transmission channel before reaching the receiver.

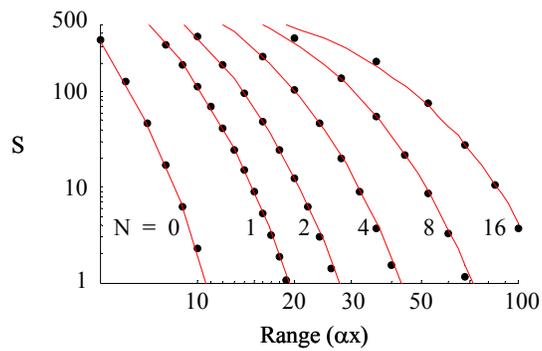

FIG. 3. Plot of the signal to noise ratio S as a function of the range $\alpha x$ in dimensionless units for a quantum cryptography system given 0, 1, 2, 4, 8, or 16 optimally spaced relays. Good agreement can be seen between the results of an exact numerical simulation (dots) and the approximate analytic results presented in the text (lines). The results shown here correspond to a quantum relay efficiency of ½ and a detector dark count probability of $10^{-5}$.

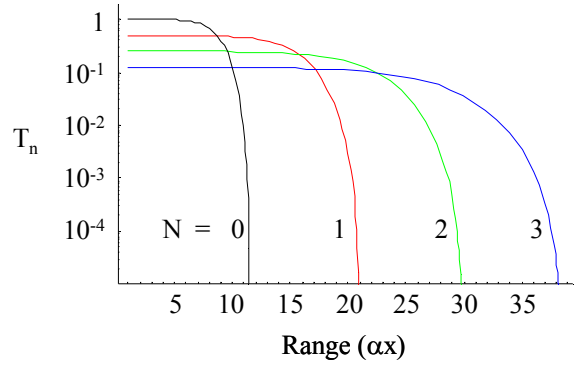

FIG. 4. Plot of the normalized throughput $T_n$ as a function of the range in dimensionless units ($\alpha x$). $T_n$ is defined as the total throughput divided by the signal rate after attenuation (prior to error correction or privacy amplification). The rapid drop in the efficiency of error correction and privacy amplification can be seen in all cases, but the use of quantum relays can extend the maximum range at which this occurs. The total throughput still decreases exponentially, however. The results shown here correspond to a quantum relay efficiency of ½ and a detector dark count probability of $10^{-5}$.